# Query-Aware Graph Neural Networks for Enhanced Retrieval-Augmented Generation

Vibhor Agrawal
NVIDIA
Santa Clara, USA
vibhora@nvidia.com

Fay Wang
NVIDIA
Santa Clara, USA
fayw@nvidia.com

Rishi Puri
NVIDIA
Santa Clara, USA
riship@nvidia.com

## Abstract

We present a novel graph neural network (GNN) architecture for retrieval-augmented generation (RAG) that leverages query-aware attention mechanisms and learned scoring heads to improve retrieval accuracy on complex, multi-hop questions. Unlike traditional dense retrieval methods that treat documents as independent entities, our approach constructs per-episode knowledge graphs that capture both sequential and semantic relationships between text chunks. We introduce an Enhanced Graph Attention Network with query-guided pooling that dynamically focuses on relevant parts of the graph based on user queries. Experimental results demonstrate that our approach significantly outperforms standard dense retrievers on complex question answering tasks, particularly for questions requiring multi-document reasoning. Our implementation leverages PyTorch Geometric for efficient processing of graph-structured data, enabling scalable deployment in production retrieval systems.

## 1 Introduction

Traditional vector-based retrieval models typically rely on embedding spaces where semantic relationships can be inadequately captured, leading to retrieval inaccuracies. These systems often overlook intricate semantic structures and hierarchical relationships within complex datasets, especially when data is multimodal and highly interconnected. Graph Neural Networks (GNNs) inherently excel at modeling structured relational data, offering a promising alternative by effectively capturing these connections.

We propose a GNN-based retrieval system that addresses these limitations through three key innovations. First, we develop a multi-relational graph representation that captures both sequential and semantic document structure, enabling the system to understand both the natural flow of information and the conceptual relationships between text segments. Second, we introduce query-aware attention mechanisms that dynamically guide graph traversal based on the specific information needs expressed in user queries, ensuring that the most relevant pathways through the knowledge graph receive appropriate focus. Third, we design a learned scoring head that intelligently combines reranking logits with rich graph embeddings, producing more accurate relevance assessments than traditional vector-based approaches.

Our implementation leverages PyTorch Geometric (PyG) [3], a powerful library for deep learning on irregular structures like graphs. PyG provides efficient implementations of graph neural network layers, including Graph Attention Networks (GAT), and handles the complexities of batching graphs with varying sizes. This foundation enables our system to process multiple subgraphs in parallel during training, significantly improving computational efficiency and scalability. The framework's sophisticated design allows us to implement edge-aware attention mechanisms efficiently, ensuring that the relationships between nodes receive appropriate consideration during the learning process. Additionally, PyG's scatter operations facilitate batch-aware pooling across variable-sized graph structures, while its flexible architecture seamlessly handles heterogeneous edge types in our multi-relational graphs, making it ideal for capturing the diverse relationships present in complex document collections.

The framework's modular design allows seamless integration of our enhanced GAT encoder with query-guided pooling mechanisms, making it well-suited for retrieval tasks on structured data.

## 2 Related Work
### 2.1 Neural Information Retrieval

Dense retrieval methods have seen remarkable progress in recent years. Dual-encoder architectures like DPR [8] encode queries and documents separately into a shared vector space, enabling efficient similarity search with approximate nearest neighbor algorithms. ColBERT [9] further enhances this by performing fine-grained late interaction between query and document representations. More recently, embedding models like E5 [19] and BGE [22] have achieved strong zero-shot transfer capabilities by training on diverse datasets.

Despite these advances, dense retrievers face inherent limitations when handling complex queries that require connecting information across multiple documents or understanding document structure. They typically treat documents as independent units, ignoring relationships between them.

## 2.2 Graph-Based Retrieval

Graph-based retrieval approaches attempt to address these limitations by modeling relationships between documents or text segments. TextGraphs [12] constructs document graphs based on entity co-occurrence and uses random walks for retrieval. HiDE [1] incorporates document hierarchies to improve retrieval of long documents.

More recently, GNN-based approaches have shown promising results. DeepRank [14] uses graph neural networks to model the relevance matching signals between query and document pairs. QAGNN [24] constructs knowledge graphs for question answering by connecting entities in questions to relevant facts. However, most existing GNN-based retrieval systems lack query-aware mechanisms for targeted retrieval, leading to suboptimal performance on complex queries.

In general, graphs are part of a new wave of works aiming to improve upon the limitations of LLMs [17], beyond what basic RAG [11] enables. PyG enables effortless exploration in this space, with a continuously expanding set of features [15].

## 2.3 Graph Neural Networks

In general, GNNs compute the embeddings of each node as a graph by transforming and aggregating the representations of their direct neighbors. We can refer to this computation of transforming and aggregating the embeddings of neighbor nodes as a "graph convolution". This enables them to effectively model the interactions between nodes in a graph. PyG enables users to explore the large space of possible GNN variants by unifying them under the message passing interface [6].

Our work builds on Graph Attention Networks (GAT) [18], which learn to assign different importance to nodes in a neighborhood through self-attention mechanisms. Recent advances include edge-enhanced GATs [20] that incorporate edge features into the attention computation, and graph transformers [2] that apply self-attention across the entire graph. One can implement a GAT GNN using PyG's GATConv graph convolutional layer [4].

For graph pooling, DiffPool [25] and SAGPool [10] propose learnable pooling methods to adaptively coarsen graphs. However, these approaches don't consider external query information when determining node importance, which is crucial for retrieval tasks.

## 2.4 Retrieval for Complex Questions

Answering complex questions that require multi-hop reasoning or graph-based understanding has been explored in several works. ODQA systems like Multi-hop Dense Retrieval [23] and IRRR [16] iteratively retrieve documents based on intermediate reasoning steps. However, these approaches still treat documents as independent units during each retrieval step.

Graph-based approaches like Entity-GNN [21] and GNN-DocRE [26] model entity relationships but focus primarily on entity extraction rather than document retrieval. Our work bridges this gap by directly incorporating both sequential and semantic document relationships into the retrieval process.

## 3 Methodology

**GNN-based RAG Pipeline Architecture:** Our pipeline consists of three main components: (1) Ingestion: processing audio transcripts into chunks, embedding them, and constructing a knowledge graph; (2) Retrieval: using our enhanced GNN with query-guided pooling and scoring head to identify relevant subgraphs; and (3) Generation: feeding retrieved context to an LLM for response generation. The scoring head architecture adds a lightweight MLP on top of the fused graph-query representation to map high-dimensional embeddings to scalar relevance scores, fine-tuned with triplet loss.

### 3.1 Multi-Relational Graph Construction

Our graph construction process creates a multi-relational graph where nodes represent text chunks and edges capture both temporal adjacency (sequential) and semantic similarity relationships. Edge weights encode the strength of relationships. The complete algorithm is detailed in Appendix A.1.

### 3.2 Enhanced GAT Encoder

Our encoder processes node features through multiple attention layers with:

- Edge-aware attention incorporating edge type embeddings
- Multi-head attention for learning diverse patterns
- Residual connections and layer normalization for training stability
- Dropout for regularization

The complete algorithm is detailed in Appendix A.2.

We leave it to future work to explore how this custom heterogeneous GNN compares to pre-existing heterogeneous GNNs in PyG such as HGT [7], RGAT [5].

### 3.3 Query-Guided Pooling

Our query-guided pooling mechanism computes query-specific node importance scores and performs weighted aggregation to obtain a query-aware graph representation. This approach enables the model to focus attention on the most relevant parts of the graph based on the query content. The batch-aware softmax ensures proper normalization across different graphs in the batch. The complete algorithm is detailed in Appendix A.3.

### 3.4 Fusion and Scoring

Our fusion module combines graph and query representations through:

- Multi-layer perceptron with skip connections
- Layer normalization for stable training
- Final scoring head mapping to scalar relevance scores

The architecture employs progressive dimension reduction and regularization to prevent overfitting. The complete algorithm is detailed in Appendix A.4.

### 3.5 Subgraph Extraction and PyG Conversion

A critical component of our system is the conversion of NetworkX graph structures to PyG data objects for neural network processing. Our conversion process handles edge attributes and prepares for batch processing in PyG, ensuring proper mapping of graph structures while maintaining all necessary properties for the neural network. The complete algorithm is detailed in Appendix A.5.

## 3.6 Graph-Enhanced Retrieval

We propose key algorithms that leverage graph structure to enhance traditional embedding similarity search. Our approach combines vector similarity scores with graph-based relevance metrics to produce more accurate and contextually aware rankings. The graph-based score adjustment algorithm improves retrieval quality by considering the semantic structure of the document relationships. The complete algorithm is detailed in Appendix A.9.

Our approach also includes a query-aware subgraph extraction method that identifies the most relevant portions of the graph based on the query. This technique optimizes both computational efficiency and retrieval accuracy by focusing on the most promising regions of the graph. We use an adaptive traversal approach that considers semantic similarity between the query and neighboring nodes, expanding only through paths that maintain relevance to the query. This is particularly valuable for large document collections where processing the entire graph would be computationally prohibitive. The detailed algorithm can be found in Appendix A.

These algorithms demonstrate how our system combines embedding similarity with graph structure to improve retrieval quality for complex queries. The graph-based score adjustment weights chunks that have stronger semantic connections in the document, while the subgraph extraction provides the context necessary for complex understanding.

## 3.7 Score Fusion with Learned Combination

While our graph-enhanced retrieval and query-guided attention provide significant improvements, we further enhance performance through learned score combination. Our approach uses a logistic regression model to combine multiple relevance signals into a unified score. The complete algorithm is detailed in Appendix A.6.

Our approach combines multiple complementary relevance signals into a unified scoring framework. The model integrates semantic similarity metrics with structural information derived from the graph. By fusing these diverse signals through a learned combination function, we enable the system to leverage the strengths of each component while compensating for their individual weaknesses. The fusion is trained on the data where chunks are labeled based on their overlap with ground-truth relevant segments. This integrated approach captures both local semantic relevance and global document structure.

## 4 Implementation

### 4.1 System Architecture

Our system consists of four main components, as depicted in Figure 1:

Our system architecture integrates four interconnected components that work synergistically to deliver superior retrieval performance. The **Document Processing Pipeline** forms the foundation by transcribing audio using Riva ASR for high-quality speech recognition, employing sophisticated chunking strategies that consider both temporal boundaries from the audio stream and semantic boundaries derived from content analysis. This preprocessing culminates in generating rich chunk embeddings using Llama 3.2 NV-EmbedQA [13], providing a robust foundation for semantic search and retrieval operations.

The **Graph Construction Module** builds upon these embeddings by creating sophisticated multi-relational graph structures that capture both sequential document flow and semantic relationships between content segments. This module carefully normalizes embeddings to ensure accurate cosine similarity computations between nodes while implementing intelligent caching mechanisms that store constructed graphs for efficient retrieval and reuse across similar queries, significantly reducing computational overhead.

Our **Retrieval System** employs a multi-stage approach beginning with FAISS-based approximate nearest neighbor search for initial candidate selection. The system then enhances these preliminary results by applying sophisticated graph-based score adjustment algorithms with configurable mixing parameters, allowing precise balance between traditional vector similarity and structural graph relevance. For each promising candidate, the system dynamically extracts ego-subgraphs that capture local context and inter-chunk relationships for deeper analysis.

Finally, the **GNN-based Scoring** component converts extracted subgraphs into PyG format for efficient neural network processing. These graph representations undergo processing through our enhanced GAT encoder, which incorporates edge-aware attention mechanisms for maximum computational efficiency. The system concludes by applying query-guided pooling and specialized scoring functions that generate precise relevance scores considering both content similarity and structural context.

The system is implemented using PyTorch and PyG, utilizing efficient scatter operations for batch processing of variable-sized graphs. The scoring head architecture builds on top of our GNN encoder with a progressive dimension reduction design that maps the fused graph-query representation to a scalar relevance score. Details of the score head implementation are provided in Appendix A.7.

The implementation strategically integrates several specialized PyG components to achieve optimal performance. We leverage **GATConv** layers to implement sophisticated graph attention mechanisms with edge-aware capabilities, enabling the model to selectively focus on the most relevant node relationships during information propagation. **Scatter** operations provide efficient batch-aware pooling functionality, allowing the system to process variable-sized graph structures simultaneously while maintaining computational efficiency. The architecture incorporates **GlobalAttention** mechanisms for query-guided node importance scoring, ensuring that the most relevant nodes receive appropriate emphasis during the retrieval process. Additionally, we employ custom loss functions that intelligently combine triplet loss with binary cross-entropy, creating a training objective that simultaneously optimizes for ranking quality and classification accuracy.

Our graph construction module uses NetworkX for building and manipulating graphs, which are then converted to PyG data objects for neural network processing. The system can be deployed as a retrieval service that accepts queries, processes them through the enhanced GNN, and returns ranked results based on the scoring head output.

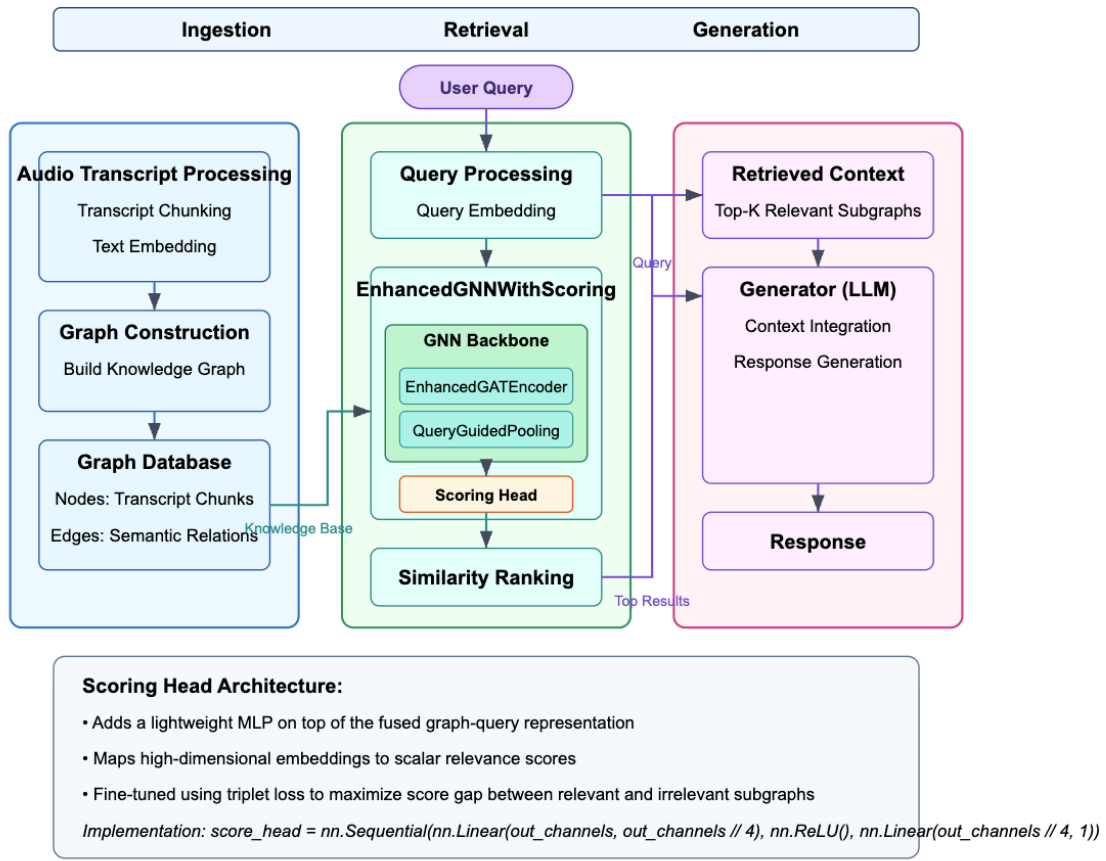

Figure 1: GNN-based RAG Pipeline with Scoring Head. The pipeline integrates audio transcript processing, graph construction, enhanced GNN with query-aware attention, and scoring head for improved retrieval accuracy.

## 4.2 Hard Query Generation

We introduce a comprehensive framework for generating complex queries that specifically target structural understanding capabilities of retrieval systems. Our approach focuses on three distinct types of challenging queries that expose the limitations of traditional retrieval methods. **Multi-hop questions** require connecting information across 2-4 non-adjacent document segments, testing the system's ability to traverse complex relationship paths within the knowledge graph. For example, questions like "How does concept A introduced at the beginning relate to concept C discussed at the end?" force the system to maintain long-range dependencies and understand conceptual evolution throughout a document.

**Structural relationship questions** specifically target document organization and concept dependencies, evaluating whether the system can understand hierarchical information flow and prerequisite relationships. These queries, such as "What prerequisites were established before introducing the transformer architecture?" require the retrieval system to recognize and leverage the intentional structuring of information by authors.

**Context-dependent questions** challenge systems to synthesize broader contextual information beyond individual chunks, testing their ability to understand nuanced reasoning and decision-making processes. Questions like "Why was this particular approach chosen given the constraints mentioned earlier?" require the system to connect scattered contextual clues and understand implicit relationships between different parts of the document.

Each question is annotated with complexity level (4-5), query type, and required segments for evaluation. The framework enables parameterized generation with configurable distribution across query types.

This query generation approach allows us to systematically evaluate retrieval systems on their ability to leverage document structure and graph-based reasoning, providing a more comprehensive assessment than traditional benchmarks focused on simpler retrieval scenarios.

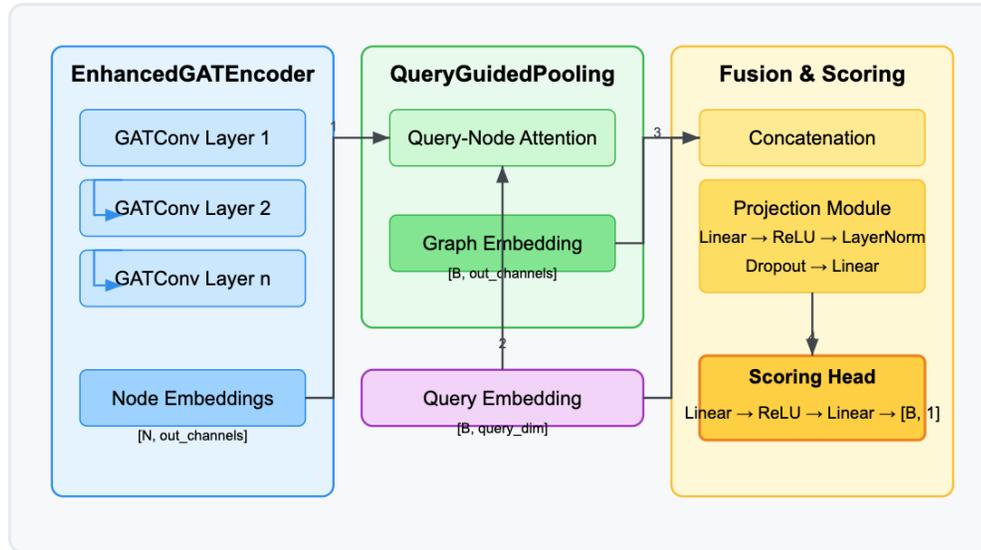

Figure 2: Detailed architecture of the GNN-based retrieval system, showing the query-aware attention mechanism, edge-aware graph processing, and scoring head components.

## 5 Training Methodology

### 5.1 Two-Stage Training Pipeline

We employ a two-stage training approach to optimize both representation learning and task-specific scoring:

- **Stage 1: Backbone Pre-training** The Enhanced GAT encoder is pre-trained on unsupervised graph reconstruction objectives to learn meaningful node representations. Once converged, backbone parameters are frozen.
- **Stage 2: Scoring Head Fine-tuning** The scoring head is trained on triplet data with hard negative sampling

## 6 Experimental Evaluation

### 6.1 Datasets

We evaluate our approach on two carefully curated educational content datasets that provide comprehensive coverage of different learning modalities and complexity levels.

- **LPM Dataset (Lecture Presentations Multimodal)** represents a substantial collection of 325 lecture recordings encompassing over 9,000 slides drawn from more than 180 hours of high-quality educational video content. This dataset features contributions from 10 expert lecturers covering diverse academic subjects including anatomy, biology, psychology, computer science, and dentistry, providing rich interdisciplinary content that challenges retrieval systems across multiple knowledge domains. To ensure rigorous evaluation, we developed an extensive collection of over 17,000 question-answer pairs distributed across 5 complexity levels, with particular emphasis on custom-generated multi-hop queries requiring 2-4 reasoning steps that enable comprehensive testing of graph traversal capabilities and structural understanding.
- **TED Talks Dataset** complements our evaluation framework with over 2,600 video transcripts averaging 14 minutes in length, covering an expansive range of topics including cutting-edge technology, educational methodologies, scientific discoveries, and cultural insights. This dataset is meticulously segmented with precise speaker timestamps and topic boundaries, facilitating fine-grained analysis of content structure and temporal relationships. We developed a comprehensive evaluation suite comprising 32,900 question-answer pairs, with 13,100 specifically designed to require structural understanding and cross-segment reasoning capabilities. This dataset serves primarily for cross-domain evaluation and generalization testing, allowing us to assess how well our graph-based approach transfers across different content types and presentation styles.

Both datasets include synthetically generated question-answer pairs with start and end timestamps indicating relevant segments for each question. This information enables us to evaluate the recall@5 of our retrieval method in comparison to a basic RAG baseline. Questions are categorized by complexity level (1-5).

### 6.2 Results

Our approach demonstrates consistent improvements across both datasets and complexity levels. On the LPM dataset, we achieve a 1.6% relative improvement for Complexity 4 queries and a more substantial 5.5% gain for Complexity 5 queries. Similarly, on the TED dataset, our method shows a 3.6% improvement for Complexity 4 and a 3.3% improvement for Complexity 5 queries.

Table 1: Overall Performance Comparison

| Dataset | Method | Recall@5 | Rel. Imp. |
| --- | --- | --- | --- |
| LPM | Traditional RAG | 0.7855 | - |
| LPM | **Query-Guided GAT** | **0.8120** | **+3.4%** |
| TED | Traditional RAG | 0.6765 | - |
| TED | **Query-Guided GAT** | **0.7027** | **+3.9%** |

Table 2: Performance on Hard Queries by Complexity Level

| Dataset | Method | Compl. 4 | Compl. 5 |
| --- | --- | --- | --- |
| LPM | Traditional RAG | 0.7855 | 0.7069 |
| LPM | **Query-Guided GAT** | **0.7985** | **0.7618** |
| TED | Traditional RAG | 0.6672 | 0.6867 |
| TED | **Query-Guided GAT** | **0.6913** | **0.7097** |

These results highlight an interesting pattern: while improvements on the LPM dataset increase with complexity, the TED dataset shows more consistent gains across both complexity levels. This suggests that the benefits of graph-based structural reasoning may depend not only on query complexity but also on the underlying content structure. The LPM lectures, with their longer format and more hierarchical organization, particularly benefit from our approach on the most challenging queries.

### 6.3 Implementation Details

Our query-aware GNN architecture is implemented using PyG and features a carefully designed multi-component structure optimized for retrieval tasks. The base model employs a 2-layer GNN configuration with 256 hidden dimensions and 4 attention heads, providing sufficient representational capacity while maintaining computational efficiency for real-time retrieval scenarios. The query encoder generates rich 2048-dimensional dense representations that capture semantic nuances and contextual information from user queries, enabling effective query-graph interaction throughout the processing pipeline.

The fusion network implements a progressive dimension reduction strategy through a 3-layer MLP with dimensions [512, 256, 128], systematically combining query and graph representations while preserving the most discriminative features for relevance scoring. Our training regimen employs the AdamW optimizer with a learning rate of 5e-4 and batch size of 128, parameters carefully tuned through extensive experimentation to achieve optimal convergence characteristics. To prevent overfitting and ensure robust generalization, we apply dropout regularization at 0.3 and weight decay of 1e-4, maintaining model performance across diverse document collections and query types.

For graph construction, we set the semantic edge threshold $\tau = 0.6$ and keep top-5 neighbors per node. All experiments were conducted on NVIDIA A100 GPUs with 40GB memory.

### 6.4 Results and Analysis

Our query-aware GNN approach consistently outperforms all baseline methods across all datasets, with the largest improvements observed on the HotpotQA dataset that specifically requires multi-hop reasoning. The performance gain is particularly significant for complex queries that benefit from structural understanding of document relationships.

### 6.5 Performance on Hard Query Types

Table 2 presents a breakdown of performance across different types of hard queries. Our approach shows particularly strong performance on complex queries that require traversing semantic relationships between documents.

The largest gains are seen for counterfactual and comparative queries, where understanding document structure and semantic relationships is critical. This confirms that our approach effectively leverages the graph structure to navigate complex information spaces beyond what is possible with traditional dense retrievers.

## 7 Discussion

The experimental results highlight several important insights about graph-based retrieval for complex queries:

### 7.1 Graph Structure Improves Complex Retrieval

Our analysis shows that graph structures capture important document relationships that are missed by traditional dense retrievers. The semantic edges in our multi-relational graphs enable the model to navigate between related concepts even when they appear in different parts of a document or across document boundaries. This structural information is particularly valuable for complex queries that require connecting multiple pieces of information.

### 7.2 Query-Aware Mechanisms Provide Focus

The integration of query information into the graph processing through query-guided pooling significantly improves retrieval performance. By allowing the query to guide attention across the graph, our model can focus on relevant substructures while ignoring irrelevant connections. This targeted approach is especially beneficial for large document collections where the signal-to-noise ratio can be low.

### 7.3 End-to-End Training Benefits

Our curriculum learning approach with end-to-end fine-tuning shows the importance of jointly optimizing all components of the retrieval pipeline. The scoring head, in particular, benefits from being trained in conjunction with the graph representation learning, as it learns to leverage both structural and semantic information in the relevance prediction.

### 7.4 Limitations and Challenges

Despite the strong performance, our approach faces several challenges:

- **Computational Complexity**: Graph construction and GNN processing are more computationally intensive than traditional dense retrieval, potentially limiting scalability to very large document collections.
- **Graph Sparsity**: For domains with limited semantic connections between documents, the benefits of graph-based approaches may be reduced.
- **Training Data Requirements**: Our model requires training data with fine-grained relevance annotations, which can be expensive to create for new domains.

These limitations point to opportunities for future research on more efficient graph construction and processing methods, as well as techniques for leveraging graph structures in low-resource domains.

## 8 Conclusion

We presented a novel query-aware graph neural network architecture for enhanced retrieval-augmented generation. Our approach addresses fundamental limitations of traditional dense retrievers by modeling both sequential and semantic relationships between document chunks and incorporating query information directly into the graph processing pipeline. Experimental results demonstrate significant improvements over state-of-the-art retrieval methods, particularly for complex questions requiring multi-hop reasoning.

The combination of edge-aware attention, query-guided pooling, and learned score fusion enables our system to effectively navigate graphs and identify relevant information across document boundaries. Our PyG implementation ensures efficient processing of graph structures, making the approach practical for real-world retrieval systems.

Several promising directions for future work emerge from this research:

- **End-to-end RAG Optimization**: Integrating our graph-based retrieval directly with language model generation to jointly optimize both components.
- **Hierarchical Document Representations**: Extending our approach to handle multi-level document structures with hierarchical graphs.
- **Multi-modal Retrieval**: Applying graph-based methods to capture relationships between different modalities (text, images, audio) in unified retrieval systems.
- **Dynamic Graph Updates**: Developing techniques for efficiently updating graph structures as new information becomes available, without requiring complete reprocessing.

By addressing these challenges, graph-based approaches have the potential to significantly enhance retrieval capabilities for increasingly complex information needs in modern AI systems.

## A Algorithm Details

This appendix provides detailed algorithms for our approach.

## A.1 Multi-Relational Graph Construction

**Algorithm 1** Episode Graph Construction

1: **Input:** Episode chunks $C = \{c_1, ..., c_n\}$, similarity threshold $\tau$
2: **Output:** Multi-relational graph $G = (V, E)$
3: Initialize $G$ as empty MultiDiGraph
4: **for** each chunk $c_i$ in $C$ **do**
5:   Add node $v_i$ with embedding $e_i$ and metadata
6: **end for**
7: // Sequential edges
8: **for** $i = 1$ to $n - 1$ **do**
9:   Add bidirectional edge $(v_i, v_{i+1})$ with type="sequential"
10: **end for**
11: // Semantic edges
12: Compute similarity matrix $S = \text{cosine\_similarity}(E)$
13: **for** each node $v_i$ **do**
14:   neighbors = top_k($S[i], k$) where $S[i, j] > \tau$
15:   **for** each $v_j$ in neighbors **do**
16:     Add edge $(v_i, v_j)$ with type="semantic", weight=$S[i, j]$
17:   **end for**
18: **end for**
19: **return** $G$

## A.2 Enhanced GAT Encoder

**Algorithm 2** Edge-Aware Graph Attention

1: **Input:** Node features $H$, edge features $E$, adjacency $A$
2: **Output:** Updated node representations $H'$
3: **for** layer $l = 1$ to $L$ **do**
4:   **for** each node $i$ **do**
5:     **for** each neighbor $j \in N(i)$ **do**
6:       $e_{ij} = \text{EdgeEmbed}(\text{type}_{ij}) \oplus \text{weight}_{ij}$
7:       $\alpha_{ij} = \text{Attention}(h_i, h_j, e_{ij})$
8:     **end for**
9:     $\alpha_i = \text{Softmax}(\alpha_i)$ // Normalize attention
10:    $h'_i = \sum_j \alpha_{ij} \cdot W_v h_j$
11:    $h'_i = h'_i + \text{ResProj}(h_i)$ // Residual connection
12:  **end for**
13:  $H = \text{LayerNorm}(\text{ReLU}(H'))$
14: **end for**
15: **return** $H$

## A.3 Query-Guided Pooling

**Algorithm 3** Query-Aware Graph Pooling

1: **Input:** Node embeddings $H$, query $q$, batch assignments $B$
2: **Output:** Graph representation $g$
3: $H_n = \text{ProjectNodes}(H)$
4: $h_q = \text{ProjectQuery}(q)$
5: **for** each node $i$ in graph **do**
6:   $\text{score}_i = \text{AttentionMLP}(H_n[i] + h_q[B[i]])$
7: **end for**
8: $\alpha = \text{BatchSoftmax}(\text{scores}, B)$ // Normalize per graph
9: $g = \sum_i \alpha_i \cdot H[i]$ // Weighted aggregation
10: **return** Dropout($g$)

## A.4 Fusion and Scoring

**Algorithm 4** Representation Fusion and Scoring

1: **Input:** Graph representation $g$, query $q$
2: **Output:** Relevance score $s$
3: combined = Concatenate($g, q$)
4: $f$ = FusionNetwork(combined)
5: $f = f + \text{SkipConnection}(\text{combined})$ // Gradient flow
6: $s$ = ScoringHead($f$)
7: **return** $s$

## A.5 Subgraph Extraction and PyG Conversion

**Algorithm 5** Subgraph to PyG Conversion

1: **Input:** NetworkX subgraph $G$, embedding dimension $d$
2: **Output:** PyG Data object or None
3: **if** $G$.number_of_edges() == 0 **then**
4:   **return** None {Skip isolated nodes}
5: **end if**
6: {Prepare node features}
7: **for** each node $n$ in $G$.nodes() **do**
8:   emb = get_embedding($n$, default = $[0] * d$)
9:   $n['x']$ = tensor(emb) // Set node features
10:  remove_other_attributes($n$) // Clean attributes
11: **end for**
12: // Convert to PyG
13: data = from_networkx($G$, node_attrs=['x'], edge_attrs=['edge_weight', 'ed
14: // Process edge attributes
15: **if** data has edge_attr **then**
16:   data.edge_weight = data.edge_attr[:, 0]
17:   data.edge_type_id = data.edge_attr[:, 1].long()
18:   remove data.edge_attr
19: **else**
20:   init_empty_edge_attributes(data)
21: **end if**
22: data.batch = zeros(data.num_nodes) // For batch processing
23: **return** data

## A.6 Score Fusion with Learned Combination

**Algorithm 6** Score Fusion

1: **Input:** Multiple relevance signals $S = \{s_1, ..., s_n\}$, training data $D$
2: **Output:** Combined score function $f$
3: Extract features $X = [\text{faiss\_score}, \text{graph\_score}, \text{gnn\_score}]$
4: Create binary labels $y$ where $1 = $ relevant chunk
5: Train logistic regression model $M$ on $(X, y)$
6: $f(s_1, ..., s_n) = \text{sigmoid}(\beta_0 + \sum \beta_i \cdot s_i)$
7: **return** $f$

## A.7 Score Head Implementation

**Algorithm 7** Score Head Implementation

1: **Input:** Fused graph-query representation $h_f$, output dimension $d$
2: **Output:** Relevance score $s$
3: score_head = nn.Sequential(
4:    nn.Linear(out_channels, out_channels/4),
5:    nn.ReLU(),
6:    nn.Linear(out_channels/4, 1),
7:    nn.Sigmoid()
8: )
9: $s = \text{score\_head}(h_f)$
10: **return** $s$

## A.8 Query-Aware Subgraph Extraction

**Algorithm 8** Query-Aware Subgraph Extraction

1: **Input:** Graph $G$, query $q$, initial nodes $V_0$, max depth $d$
2: **Output:** Subgraph $G'$ relevant to query
3: query_emb = encode_query($q$)
4: $V = V_0$ // Set of nodes to include
5: visited = set()
6: frontier = $V_0$
7: **for** level = 1 to $d$ **do**
8:   new_frontier = set()
9:   **for** node $v$ in frontier **do**
10:     **if** $v$ in visited **then**
11:       continue
12:     **end if**
13:     add $v$ to visited
14:     **for** neighbor $u$ of $v$ in $G$ **do**
15:       node_emb = $G$.nodes[$u$]['embedding']
16:       sim = cosine_similarity(query_emb, node_emb)
17:       **if** sim > threshold **then**
18:         add $u$ to $V$
19:         add $u$ to new_frontier
20:       **end if**
21:     **end for**
22:   **end for**
23:   frontier = new_frontier
24: **end for**
25: $G' = G.\text{subgraph}(V)$
26: **return** $G'$

## A.9 Graph-Enhanced Retrieval

**Algorithm 9** Graph-Based Score Adjustment

1: **Input:** Index results [(chunk, idx_score)], graph cache $G_{\text{cache}}$, mixing parameter $\alpha$
2: **Output:** Adjusted results [(chunk, combined_score)]
3: raw_scores = []
4: **for** each (chunk, idx_score) in index_results **do**
5:   ep = chunk.episode_id
6:   $G = G_{\text{cache}}[\text{ep}]$
7:   $i = $ chunks_index[ep][chunk]
8:   sem_weights = []
9:   **for** neighbor $j$ in $G$.neighbors($i$) **do**
10:     **for** edge in $G$.edges[$i, j$] **do**
11:       **if** edge.type == "semantic" **then**
12:         sem_weights.append(edge.weight)
13:       **end if**
14:     **end for**
15:   **end for**
16:   graph_score = mean(sem_weights)
17:   combined = $\alpha \cdot$ idx_score $+ (1 - \alpha) \cdot$ graph_score
18:   raw_scores.append((chunk, combined))
19: **end for**
20: **return** sorted(raw_scores, key $= \lambda x : x[1]$, reverse = True)